\documentclass[prd,twocolumn,nofootinbib,amsmath,amssymb,floatfix,superscriptaddress]{revtex4}
\usepackage{multirow}
\usepackage{graphicx}
\usepackage{epstopdf}
\usepackage{dcolumn}
\usepackage{bm}
\usepackage{threeparttable}
\usepackage{subfigure}
\usepackage{newtxtext}
\usepackage{color,txfonts}
\usepackage{ulem}
\definecolor{blue0}{rgb}{0,0,0.6}
\usepackage[colorlinks,linkcolor=blue0,anchorcolor=blue0,citecolor=blue0,urlcolor=blue0]{hyperref}

\newcommand{\beq}{\begin{equation}}
\newcommand{\eeq}{\end{equation}}
\newcommand{\beqa}{\begin{eqnarray}}
\newcommand{\eeqa}{\end{eqnarray}}

\begin{document}

\title{Constraints on dark matter annihilation from a nearby subhalo candidate}

\author{Ben-Yang Zhu}
\affiliation{Key Laboratory of Dark Matter and Space Astronomy, Purple Mountain Observatory, Chinese Academy of Sciences, Nanjing 210023, China}
\affiliation{School of Astronomy and Space Science, University of Science and Technology of China, Hefei 230026, China}
\author{Yun-Feng Liang}
\email{liangyf@gxu.edu.cn}
\affiliation{Guangxi Key Laboratory for Relativistic Astrophysics, School of Physics Science and Technology, Guangxi University, Nanning 530004, China}
\author{Xiaoyuan Huang}
\email{xyhuang@pmo.ac.cn}
\affiliation{Key Laboratory of Dark Matter and Space Astronomy, Purple Mountain Observatory, Chinese Academy of Sciences, Nanjing 210023, China}
\affiliation{School of Astronomy and Space Science, University of Science and Technology of China, Hefei 230026, China}

\date{\today}

\begin{abstract}

A recent analysis of pulsar timing data has reported evidence for a massive ($\sim 6 \times 10^7 M_{\odot}$) dark matter subhalo located only $\sim 0.8$ kpc from Earth. This candidate implies an exceptionally large $J$-factor of $\sim 10^{23}\,{\rm GeV^2\,cm^{-5}}$, exceeding that of known classical dwarf spheroidal galaxies by orders of magnitude and rivaling the Galactic Center. In this work, we utilize more than 17 years of \textit{Fermi}-LAT data to search for gamma-ray emission from this subhalo. We identify a tentative excess in the region with ambiguous origin. Adopting a conservative strategy, we retain this excess without modeling additional astrophysical components, treating it instead as unmodeled background to derive upper limits on the dark matter annihilation cross-section for the $b\bar{b}$ and $\tau^+\tau^-$ channels. Despite this conservative treatment, the resulting limits remain stringent due to the exceptionally  large $J$-factor. Subject to the dynamical confirmation of the subhalo, these constraints are potentially orders of magnitude stronger than those obtained from combined analyses of dwarf spheroidal galaxies and blind subhalo searches.
\end{abstract}

\maketitle

\section{Introduction}
According to the standard lambda cold dark matter ($\Lambda$CDM) cosmological model, dark matter (DM) constitutes $\sim$27\% of the current energy density of the Universe \cite{Planck:2018vyg}. In this paradigm, structure formation proceeds in a bottom-up fashion, where small halos form first and subsequently merge to build larger systems. High-resolution $N$-body simulations, such as Via Lactea II~\cite{Diemand:2008in} and Aquarius~\cite{Springel:2008cc}, confirm that this hierarchical assembly, combined with incomplete tidal disruption, results in a vast population of surviving DM subhalos orbiting within massive galactic halos like that of the Milky Way~\cite{Moore:1999nt}. While the massive subhalos ($\gtrsim 10^8\,M_{\odot}$) are typically associated with dwarf spheroidal galaxies (dSphs) which are rich in stars and gas and serve as robust targets for DM searches~\cite{Tsai:2012cs, Fermi-LAT:2015att,Fermi-LAT:2016uux,Geringer-Sameth:2014qqa,Li:2015kag,Hoof:2018hyn,Li:2021vqg, Guo:2022rqq,McDaniel:2023bju,LHAASO:2024upb}, theoretical models predict a much larger population of low-mass subhalos (below $10^8 M_{\odot}$) \cite{Sanchez-Conde:2013yxa,Moline:2016pbm}. These dark subhalos are expected to be essentially baryon-free and thus elude detection at optical wavelengths. Identifying such dark substructures is crucial for testing the small-scale predictions of $\Lambda$CDM, yet their lack of starlight makes them invisible to optical surveys.

If DM is composed of weakly interacting massive particles (WIMPs), these subhalos could produce detectable gamma-ray signals through particle annihilation or decay~\cite{Kuhlen:2008aw,Baltz:2008wd,Anderson:2010df}. 
Over the past decade, the Fermi Large Area Telescope (\textit{Fermi}-LAT) \cite{Fermi-LAT:2009ihh} has been the premier instrument for searching for such signatures. Dedicated searches for dark satellites using early \textit{Fermi}-LAT catalogs~\cite{Fermi-LAT:2012fij}, as well as numerous follow-up studies of unassociated sources in the 2FGL, 3FGL and 4FGL catalogs~\cite{Zechlin:2012by,Berlin:2013dva, Bertoni:2015mla,Schoonenberg:2016aml,Calore:2016ogv,Hooper:2016cld,Coronado-Blazquez:2019puc}, have identified a handful of spatially extended or spectrally curved candidates, such as 3FGL J2212.5+0703 and 3FGL J1924.8–103, whose spectra appeared compatible with DM annihilation~\cite{Bertoni:2016hoh,Wang:2016xjx,Xia:2016uog}. However, distinguishing DM subhalo candidates from conventional astrophysical sources remains a significant challenge. With improved statistics and refined analysis techniques, such as modeling spectral curvature and cross-matching with \textit{Gaia} stellar counterparts, the majority of previously proposed candidates have been reclassified as pulsars or active galactic nuclei (AGN), or have lost statistical significance~\cite{Coronado-Blazquez:2019pny}. More recently, population-level analyses have modeled the entire unassociated source population as a mixture of astrophysical sources and DM subhalos determined by Monte Carlo simulations, again finding no evidence for a DM contribution and instead deriving upper limits on the annihilation cross section~\cite{2025arXiv250314584A}. 
In addition, searches for gamma-ray spectral lines from subhalos also did not find definitive evidence so far~\cite{Liang:2017vmq,Li:2018xzs}. A fundamental limitation of these blind sky searches is the lack of independent dynamical constraints. Lacking kinematic data, the distance, mass, and consequently the expected DM signal intensity ($J$-factor) of a candidate source remain degenerate.

Detecting DM subhalos directly through their gravitational influence breaks this degeneracy. Recently, Chakrabarti et al.~(2025)~\cite{2025arXiv250716932C} introduced a novel method based on pulsar timing measurements as essentially Galactic accelerometers. By analyzing the acceleration field of binary and solitary pulsars within the Milky Way, they identified a localized perturbation of the gravitational potential in the vicinity of the Solar System. Their dynamical modeling favors an enhanced mass located at a Galactocentric distance of $X \approx 7.4$ kpc (corresponding to a heliocentric distance of $\sim 0.8$ kpc) with an inferred mass of approximately $10^7\,M_{\odot}$, while disfavoring interpretations in terms of known gas or stellar structures~\cite{2025arXiv250716932C}. This result provides a rare example of a well-localized dark subhalo candidate whose position and mass are inferred primarily from its gravitational imprint. Such a gravitationally selected target allows us to move from blind searches across the sky to a sensitive, targeted search at a known location.

Motivated by the discovery reported in Chakrabarti et al.~(2025)~\cite{2025arXiv250716932C}, in this work we perform a dedicated gamma-ray search from this subhalo candidate using 17 years of \textit{Fermi}-LAT Pass~8 data. Given the candidate's proximity ($\sim 0.8$ kpc) and substantial mass, its predicted astrophysical $J$-factor would exceed that of the brightest dSphs under the fiducial NFW parameters inferred in Chakrabarti et al.~\cite{2025arXiv250716932C}, offering an unprecedented opportunity for sensitivity. Assuming that the gravitational perturbation arises from a WIMP DM subhalo, we model its extended spatial emission with an NFW density profile and perform a binned likelihood analysis to search for excess gamma-ray emission at the reported position. In the absence of a significant signal, we use the dynamically constrained mass and distance, together with the large inferred $J$-factor, to place stringent constraints on the DM annihilation cross section.

The remainder of this paper is organized as follows. In Section~\ref{sec2}, we describe the dynamical properties of the nearby subhalo candidate and derive the spatial template for its expected dark matter signal. Section~\ref{sec3} details the \textit{Fermi}-LAT data selection, background modeling, and the search for residual gamma-ray emission in the region of interest. In Section~\ref{sec4}, we present the statistical framework used to constrain the dark matter annihilation cross section, incorporating both geometric and mass-related uncertainties. Finally, we summarize our findings and discuss their broader implications in Section~\ref{sec5}.

\section{J-factor of the Nearby Subhalo}
\label{sec2}

We adopt the properties of the nearby DM subhalo candidate recently inferred from the acceleration field of binary pulsars by Chakrabarti et al.~\cite{2025arXiv250716932C}. Their Bayesian analysis yielded Bayes factors of $20$--$40$, providing tentative evidence for a localized mass perturbation. Following their dynamical modeling of the perturbation using a Navarro--Frenk--White (NFW) density profile~\cite{Navarro_1997}, we utilize their best-fit subhalo virial mass of:
\begin{equation}
    M_{\rm sub} = (6.19^{+1.92}_{-2.03})\times 10^7\,M_{\odot}\,.
\end{equation}
The corresponding centroid in the Galactocentric Cartesian coordinates is:
\begin{equation}
    X = 7.47^{+0.21}_{-0.14}~{\rm kpc},\quad
    Y = 0.38^{+0.11}_{-0.16}~{\rm kpc},\quad
    Z = 0.21^{+0.06}_{-0.11}~{\rm kpc}\,.
\label{eq:xyz}
\end{equation}
Adopting a Solar position at $X_\odot \approx 8.122$~kpc~\cite{GRAVITY:2018ofz}, these coordinates correspond to a heliocentric distance of $d \simeq 0.78~{\rm kpc}$. Propagating the quoted covariance on $(X,Y,Z)$ yields a distance uncertainty of $\sigma_d \simeq 0.17~{\rm kpc}$.

The DM density distribution is modeled using the standard NFW profile~\cite{Navarro_1997}:
\begin{equation}
\label{eq:NFW}
    \rho_{\rm DM}(r) = \frac{\rho_s}{\left({r}/{r_s}\right)\left(1 + {r}/{r_s}\right)^2}\,,
\end{equation}
where $r$ is the distance from the halo center. The characteristic scale density $\rho_s$ is determined by the virial mass $M_{\rm sub}$, scale radius $r_s$ and concentration $c$ via
\begin{equation}
\label{eq:rhos}
    \rho_s = \frac{M_{\rm sub}}{4\pi r_s^3 \left[\ln(1+c) - \dfrac{c}{1+c}\right]}\,.
\end{equation}
In Chakrabarti et al.~\cite{2025arXiv250716932C}, the NFW profile fit was performed by fixing the scale radius $r_s = 0.1$ kpc and concentration $c = 30$, based on expectations from cosmological simulations for halos of this mass scale~\cite{Bullock_2017}. We adopt these parameters to derive the fiducial $J$-factor, while acknowledging that the internal density profile is not uniquely constrained by the current pulsar timing data. Crucially, the inferred mass $M_{\rm sub}$ is coupled to these geometric assumptions. Therefore, we strictly adhere to this self-consistent set of parameters rather than exploring a broader range of unconstrained profiles.

Given the subhalo's proximity ($d \sim 0.8$ kpc) and physical size ($r_s \sim 0.1$ kpc), the source would extend a large angle on the sky and cannot be approximated as a point source. Consequently, we model the expected gamma-ray emission as a spatially extended source. The surface brightness is proportional to the differential $J$-factor, defined as the line-of-sight (l.o.s.) integral of the squared DM density in a specific direction $\psi$:
\begin{equation}
\label{eq:dJdOmega}
    \frac{dJ}{d\Omega}(\psi) = \int_{\rm l.o.s.} \rho_{\rm DM}^2\big(r(l, \psi)\big) \, {\rm d}l \,.
\end{equation}
We compute this value for every pixel in our region of interest (ROI) to generate a spatial template map. The total $J$-factor is then the integral of this map over the ROI solid angle $\Delta\Omega$:
\begin{equation}
    J = \int_{\Delta\Omega} \frac{dJ}{d\Omega} \, {\rm d}\Omega \,.
\end{equation}

In standard indirect-detection likelihood analyses, $J$-factor uncertainties are often treated as nuisance parameters with a lognormal prior while the source morphology is kept fixed~\cite{Huang:2011xr,Tsai:2012cs, Fermi-LAT:2013sme,Fermi-LAT:2015att,Fermi-LAT:2016uux}. However, for a nearby subhalo, the distance uncertainty ($\sigma_d/d \sim 20\%$) significantly impacts both the signal normalization  and the apparent angular size, $\theta_s \approx r_s/d$. Varying the distance implies modifying the projected surface-brightness distribution on the sky. Thus, profiling over $J$ while fixing the spatial template would therefore be geometrically inconsistent.

To rigorously address these dependencies, we adopt a decoupled treatment that separates geometric uncertainty from the statistical uncertainty on the subhalo mass. We first consider the geometric uncertainty arising from the distance estimate. Since the heliocentric distance determines the angular scale of the emission, we generate three distinct spatial templates corresponding to the nominal distance ($d = 0.78$~kpc) and its $1\sigma$ bounds ($d-\sigma_d = 0.61$~kpc and $d+\sigma_d = 0.95$~kpc). For each distance, we compute the pixelated $dJ/d\Omega$ map and the total integrated $J$-factor over the $20^\circ\times20^\circ$ ROI, yielding:
\begin{equation}
J \simeq [0.79,1.14,1.82] \times 10^{23}\ {\rm GeV^2cm^{-5}},
\end{equation}
where the three values correspond to the three heliocentric distances. We perform three independent likelihood analyses using these templates. The analysis using the nominal distance provides our fiducial limit, while the analyses using the near and far templates define the uncertainty band arising from the subhalo's position.

Within each of these three spatial configurations, we account for the statistical uncertainty associated with the subhalo mass. Variations in mass affect the density normalization $\rho_s$ and consequently the overall scale of $J$, but leave the angular morphology essentially unchanged for a fixed distance. Since $J \propto M_{\rm sub}^2$ (holding $r_s$, $c$ and $d$ constant), the fractional uncertainty on $M_{\rm sub}$ translates directly into a log-normal uncertainty in the $J$-factor normalization. Based on the pulsar timing constraints, we derive a normalization uncertainty of $\sigma_{\log_{10}J} \simeq 0.29$\,dex. This would be incorporated as a nuisance parameter scaling the spatial template in the likelihood analysis described in Sec.~\ref{sec3}. This approach ensures that the coupling between the subhalo's distance, total brightness, and angular size is consistently modeled.

\section{data analysis }
\label{sec3}

In this work, we utilize over 17 years of \textit{Fermi}-LAT Pass~8 data, covering the Mission Elapsed Time (MET) interval 239557417--785293911 (2008-08-04 to 2025-11-20). We select \texttt{SOURCE} class events (\texttt{evclass=128}) in the energy range 500~MeV--500~GeV. To minimize contamination from the Earth's limb, we apply a maximum zenith angle cut of $100^\circ$. Standard data quality selections are imposed using \texttt{(DATA\_QUAL>0 \&\& LAT\_CONFIG==1)}. Given the subhalo scale radius $r_s = 0.1$~kpc and the distance range $d = 0.61$--$0.95$~kpc (Sec.~\ref{sec2}), the characteristic angular scale set by the NFW scale radius is $\theta_s \equiv \arctan(r_s/d) \approx r_s/d$, which varies from $\sim 6^\circ$ to $\sim 9^\circ$. We therefore adopt a $20^\circ\times20^\circ$ region of interest (ROI) centered on the best-fit subhalo position (J2000: R.A.\,$=267.78^\circ$, Decl.\,$=4.70^\circ$), which is derived according to the Galactocentric Cartesian coordinates (i.e., Eq.~(\ref{eq:xyz})). We bin the data using a spatial pixel size of 0.1$^\circ$ and 30 logarithmic energy bins between 0.5–500 GeV. The baseline background model is constructed using the 4FGL-DR4 catalog~\cite{Fermi-LAT:2022byn, ballet2024fermilargeareatelescope}, including all point-like and extended sources within $17^\circ$ of the ROI center, together with the standard Galactic diffuse emission model (\texttt{gll\_iem\_v07.fits}) and the isotropic template (\texttt{iso\_P8R3\_SOURCE\_V3\_v1.txt}) recommended by the \textit{Fermi}-LAT Collaboration.

\begin{figure}
    \centering
    \includegraphics[width=0.49\textwidth]{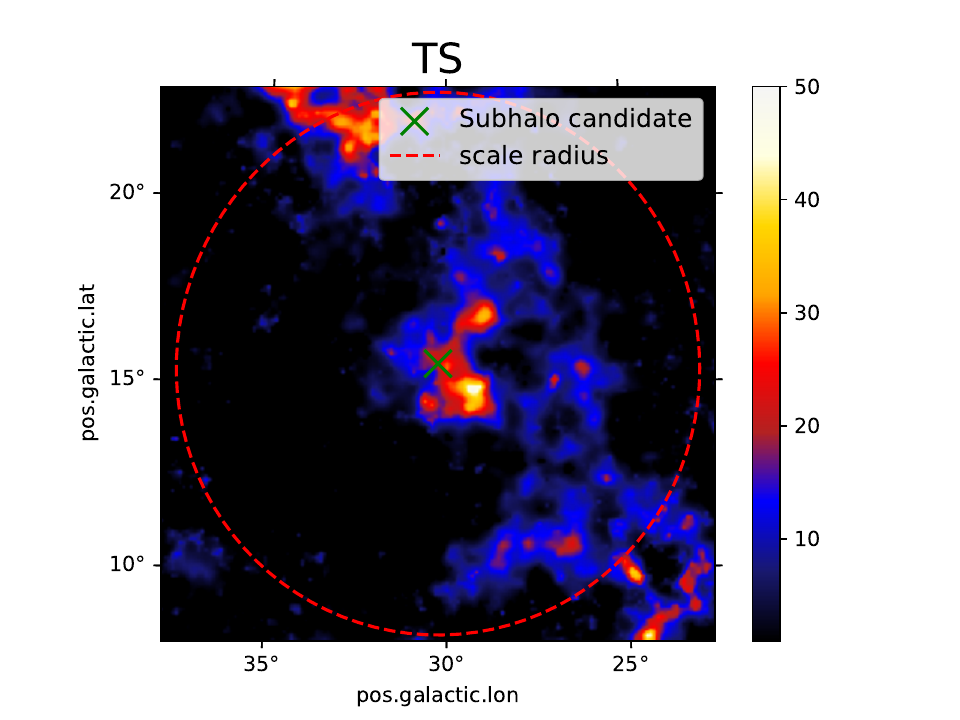}
    \caption{Residual TS map of the $15^\circ \times 15^\circ$ region centered on the nearby subhalo (marked with a green cross), the red dashed line represents the scale radius of $r_s=0.1\,{\rm kpc}$ for the best-fit distance.} 
    \label{tsmap}
\end{figure}

We perform a standard binned Poisson likelihood analysis using \texttt{Fermitools} (v2.4.0), with the P8R3\_SOURCE\_V3 instrument response functions. In the baseline fit (without an explicit DM component), the spectral parameters of all catalog sources within $8^\circ$ of the ROI center are left free to vary, as are the normalizations of the two diffuse components and of bright sources just outside the $8^\circ$ radius. The significance of additional sources or emission features is quantified using the test statistic, ${\rm TS} = -2\ln\left({\mathcal{L}_0}/{\mathcal{L}_1}\right)$,
where $\mathcal{L}_0$ and $\mathcal{L}_1$ are the maximum likelihood values under the null (background-only) and alternative hypotheses, respectively~\cite{Mattox:1996zz, ROLKE2005493}. To assess the goodness of fit and to search for unmodeled emission, we compute a residual TS map using the \texttt{gttsmap} tool. This procedure adds a putative point source with a power-law spectrum at each pixel in the ROI and evaluates the resulting TS improvement relative to the baseline model. Figure~\ref{tsmap} shows the residual TS map in a $15^\circ \times 15^\circ$ sub-region centered on the subhalo. The red dashed circle indicates the angular scale radius of the subhalo, $\theta_s \simeq 7.3^\circ$, corresponding to the nominal distance $d=0.78$~kpc. We find a best-fit residual excess with ${\rm TS}>25$ located approximately $1.5^\circ$ from the subhalo centroid (green cross).

When modeled explicitly as a point source at its TS maximum, this excess reaches ${\rm TS}\sim 50$, which would be formally sufficient to claim a detection in standard \textit{Fermi}-LAT catalog analyses. However, it is still insufficient to robustly determine its physical nature, and it could be compatible with a dark matter signal, an unresolved astrophysical source, or a local fluctuation of the Galactic diffuse emission. Given this ambiguity, we do not attempt to interpret the excess. To avoid biasing our dark matter constraints, we therefore adopt a conservative strategy in Sec.~\ref{sec4}. We do not introduce any additional astrophysical component to absorb this excess, nor do we mask the region. Instead, we retain the baseline background model and allow the DM template and background parameters to fit this residual emission within the likelihood framework. This choice naturally leads to weaker (more conservative) upper limits than an analysis that explicitly subtracts the excess, but ensures that the limits are not artificially tightened by removing a signal of uncertain origin.

\section{constraints on dark matter}
\label{sec4}

The differential gamma-ray flux expected from DM self-annihilation is given by:
\begin{equation}
    \frac{{\rm d}\Phi}{{\rm d}E\,{\rm d}\Omega}(E, \psi)
    = \frac{1}{4\pi}\,\frac{\langle\sigma v\rangle}{2 m_{\chi}^2}\,\frac{{\rm d}N}{{\rm d}E}(E)\,\frac{{\rm d}J}{{\rm d}\Omega}(\psi)\,,
\end{equation}
where $\langle\sigma v\rangle$ is the velocity-averaged annihilation cross section, $m_\chi$ is the DM particle mass, and ${\rm d}N/{\rm d}E$ is the photon spectrum per annihilation event, obtained from \texttt{PPPC4DMID}~\cite{Cirelli:2010xx} with electroweak corrections. The term ${\rm d}J/{\rm d}\Omega$ represents the spatial morphology of the subhalo as derived in Sec.~\ref{sec2}. The expected counts map used in the likelihood fit is obtained by convolving this intensity with the instrument response functions.

Following the strategy outlined in Sec.~\ref{sec2}, we treat the uncertainties in the subhalo properties in two steps. First, the distance uncertainty is handled by performing three independent likelihood analyses using the distinct spatial templates (${\rm d}J/{\rm d}\Omega$ maps) corresponding to $d-\sigma_d=0.61$~kpc, $d=0.78$~kpc, and $d+\sigma_d=0.95$~kpc. Then, within each of these spatial configurations, we incorporate the uncertainty on the subhalo mass via a log-normal prior on the $J$-factor normalization~\cite{Huang:2011xr,Tsai:2012cs, Fermi-LAT:2013sme,Fermi-LAT:2015att,Fermi-LAT:2016uux}:
\begin{align}
    \mathcal{L}_J(J \mid J_{\rm obs}, \sigma_J) = &
    \frac{1}{\ln(10)\,J \sqrt{2\pi}\,\sigma_J}\times \nonumber \\
    &\exp\left[-\frac{\big(\log_{10} J - \log_{10} J_{\rm obs}\big)^2}{2\sigma_J^2}\right],
\end{align}
where $\sigma_J = 0.29$~dex is the dispersion propagated from the mass estimate.

For a given DM mass $m_\chi$, the analysis is performed by maximizing the joint likelihood function:
\begin{equation}
   \mathcal{L}_{\rm tot}(\mu, \boldsymbol{\theta} \mid D) =
   \mathcal{L}_{\rm LAT}(\mu, \boldsymbol{\theta} \mid D) \times
   \mathcal{L}_J(J \mid J_{\rm obs}, \sigma_J)\,,
\end{equation}
where $\mu \equiv \langle \sigma v \rangle$ is the signal parameter of interest, $\boldsymbol{\theta}$ denotes the nuisance parameters (including the background spectral parameters and the $J$-factor normalization), and $D$ represents the observed \textit{Fermi}-LAT data.

We first quantify the significance of a potential DM signal by comparing the best-fit model including a DM component to the null hypothesis (no DM, $\mu=0$). Let $(\hat{\mu}, \hat{\boldsymbol{\theta}})$ denote the global maximum-likelihood estimators obtained by maximizing $\mathcal{L}_{\rm tot}(\mu, \boldsymbol{\theta} \mid D)$ over $\mu \ge 0$ and $\boldsymbol{\theta}$, and let $\hat{\boldsymbol{\theta}}_0$ denote the nuisance parameters that maximize $\mathcal{L}_{\rm tot}$ under the null hypothesis $\mu=0$. The Test Statistic of dark matter is defined as:
\begin{equation}
    {\rm TS}_{\rm dm} = -2 \ln \left[
    \frac{\mathcal{L}_{\rm tot}(0, \hat{\boldsymbol{\theta}}_0)}
         {\mathcal{L}_{\rm tot}(\hat{\mu}, \hat{\boldsymbol{\theta}})}
    \right].
\end{equation}
Our analysis yields a maximum ${\rm TS}_{\rm dm} \approx 60$ for the $b\bar{b}$ final states, and a maximum ${\rm TS}_{\rm dm} \approx 30$ for the $\tau^+\tau^-$ final states. However, we note that formal statistical significance alone does not guarantee a dark matter detection in complex regions. Notably, the Galactic center excess shows a far stronger signal but remains physically inconclusive~\cite{Goodenough:2009gk,Hooper:2010mq,Hooper:2013rwa,Gordon:2013vta,Abazajian:2014fta,Daylan:2014rsa,Zhou:2014lva,Calore:2014xka,Huang:2015rlu,Fermi-LAT:2017opo,Zhong:2019ycb,Lee:2015fea,Bartels:2015aea,Macias:2016nev,Bartels:2017vsx,Leane:2019xiy,Zhu:2022tpr,2025arXiv251115793L}. Accordingly, we focus on using these data to place constraints. To derive upper limits on the annihilation cross section, we compute the profile likelihood as a function of $\mu$:
\begin{equation}
    \lambda(\mu) = \frac{\mathcal{L}_{\rm tot}(\mu, \hat{\hat{\boldsymbol{\theta}}}_\mu)}{\mathcal{L}_{\rm tot}(\hat{\mu}, \hat{\boldsymbol{\theta}})},
\end{equation}
where $\hat{\hat{\boldsymbol{\theta}}}_\mu$ denotes the nuisance parameters that maximize the likelihood for a fixed value of $\mu$. The one-sided 95\% confidence level (CL) upper limit, $\mu^{95}$, is defined as the value of $\mu$ (where $\mu > \hat{\mu}$) for which the log-likelihood decreases by $2.71/2$ relative to the global maximum~\cite{ROLKE2005493}:
\begin{equation}
    -2 \ln \lambda(\mu^{95}) = 2.71 \,.
\end{equation}

\begin{figure*}[t]

    \centering
    \includegraphics[width=0.45\textwidth]{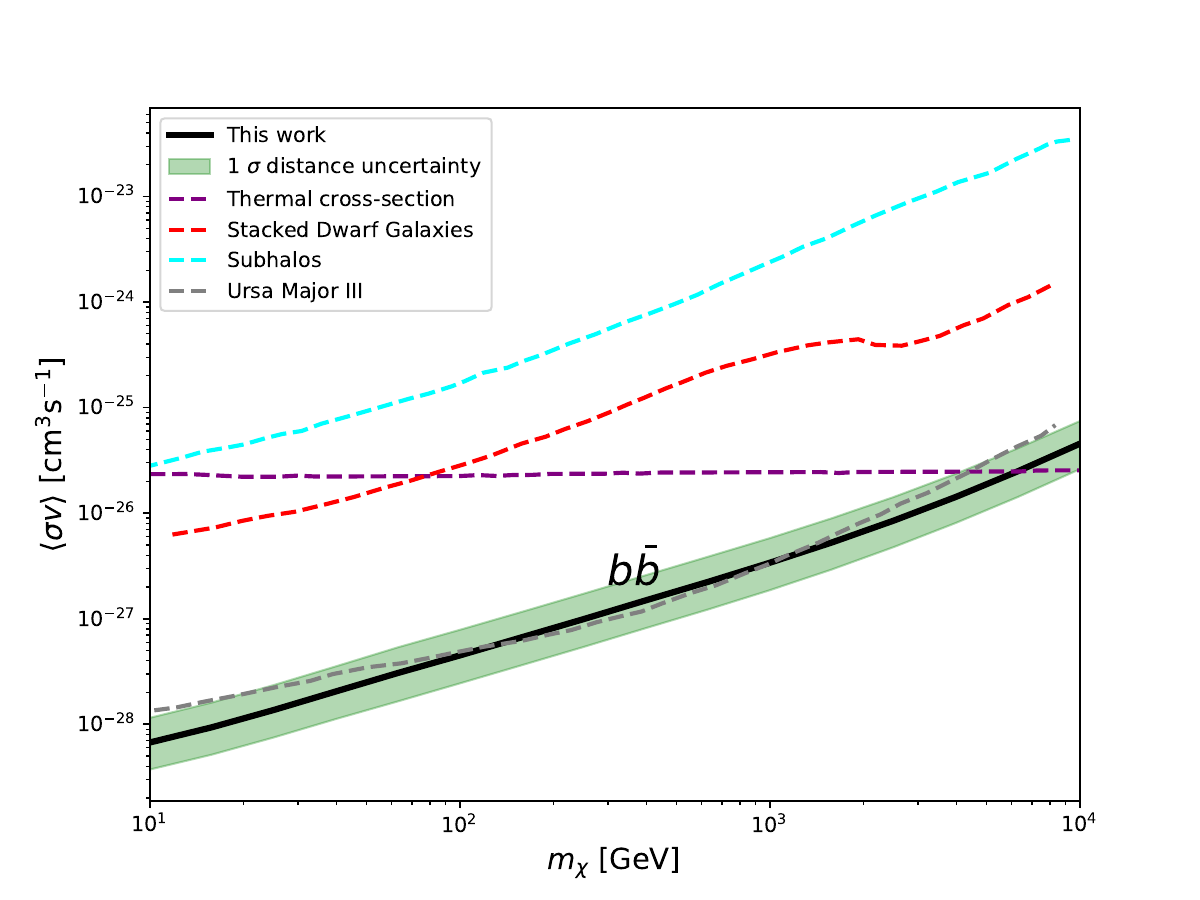}
    \includegraphics[width=0.45\textwidth]{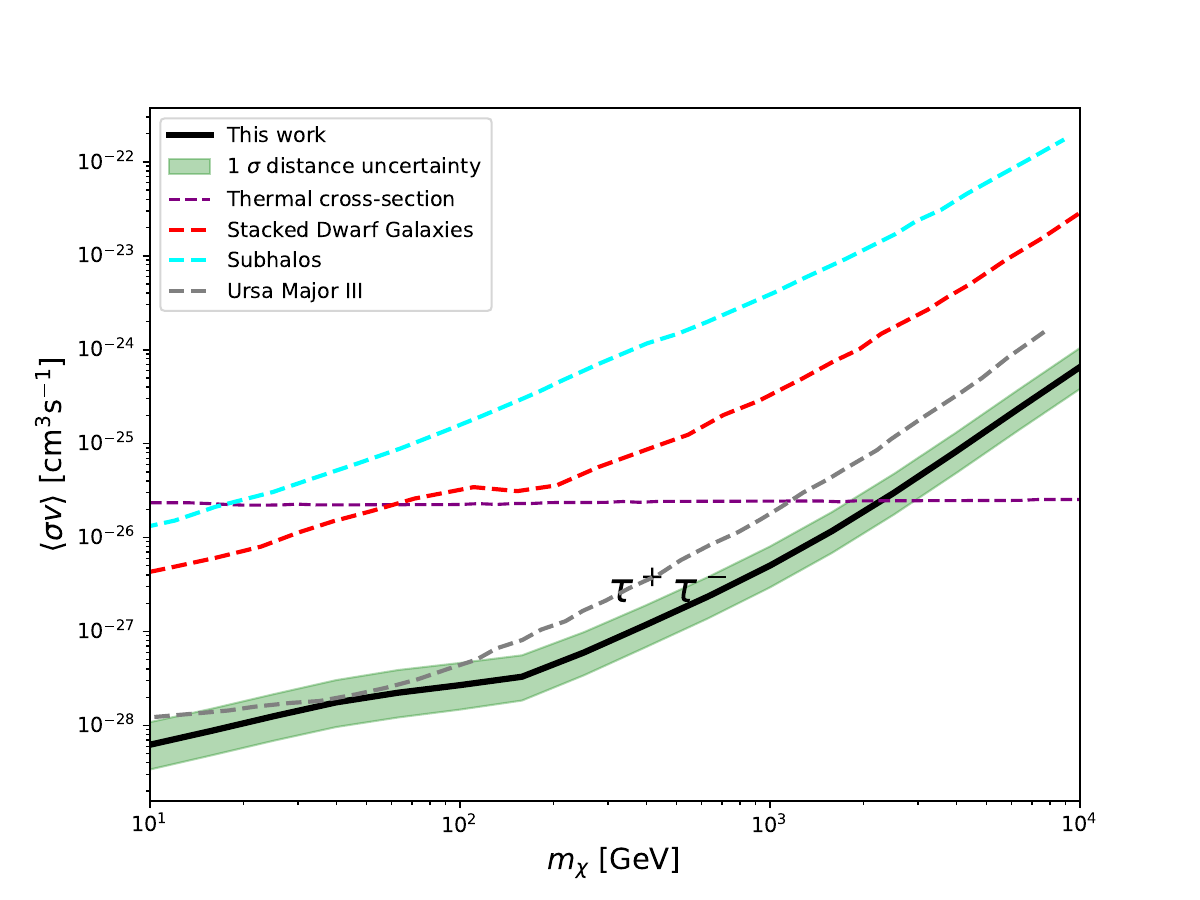}
    \caption{Constraints on the dark matter annihilation cross section $\langle \sigma v\rangle$ at 95\% CL for the $b\bar{b}$ (left) and $\tau^+\tau^-$ (right) channels. The solid black curve represents the fiducial limit derived in this work, while the green shaded band indicates the uncertainty arising from the $1\sigma$ variation in the subhalo distance. For comparison, we show limits from stacked dwarf spheroidal galaxies~\cite{McDaniel:2023bju} (red dashed line), subhalo candidates among \textit{Fermi} unidentified sources~\cite{Coronado-Blazquez:2019pny} (cyan dashed line), and Ursa Major III~\cite{Crnogorcevic:2023ijs} (grey dashed line). The canonical thermal relic cross section~\cite{Steigman:2012nb} is indicated by the purple dashed line.}
    \label{constraints}
\end{figure*}

We perform this procedure for DM masses in the range $m_\chi = 10~{\rm GeV}$--$10~{\rm TeV}$, considering annihilation into the $b\bar{b}$ and $\tau^+\tau^-$ final states. As noted above, the unmodeled residual emission in the ROI would be partially absorbed by the DM template, resulting in a positive best-fit signal strength ($\hat{\mu} > 0$) and the non-zero ${\rm TS}_{\rm dm}$ values. Deriving upper limits relative to this global maximum (rather than zero) yields conservative constraints, ensuring we do not artificially tighten the limits by subtracting a signal of uncertain origin. The resulting exclusion curves are shown in Fig.~\ref{constraints}. The solid curves correspond to the fiducial limits derived using the best-fit distance template ($d=0.78$~kpc), while the green shaded band shows the systematic uncertainty spread induced by varying the distance within the $\pm 1\sigma$ range ($0.61$--$0.95$~kpc). And these limits are potentially orders of magnitude stronger than those obtained from dwarf spheroidal galaxies~\cite{McDaniel:2023bju} and blind subhalo searches~\cite{Coronado-Blazquez:2019pny}, and are comparable to recent nominal constraints for the ultra-faint satellite Ursa Major III, which would be severely relaxed if uncertainties are considered~\cite{Crnogorcevic:2023ijs}.

\section{Conclusions \& Discussion}
\label{sec5}

$N$-body simulations predict that Milky Way-like halos should host a vast population of dark matter subhalos over a wide range of masses. While some of the most massive subhalos can be identified with dwarf spheroidal galaxies and have been extensively used as targets for indirect DM searches, lower-mass, baryon-poor subhalos have so far remained undetected electromagnetically. A recent analysis of pulsar timing data has provided dynamical evidence for a massive ($\sim 10^7\,M_\odot$) perturber located only $\sim 0.8$~kpc from the Sun. If it is a dark matter subhalo, its proximity and inferred mass imply an exceptionally large astrophysical $J$-factor, $J_{\rm ROI} \sim 10^{23}\,{\rm GeV^{2}\,cm^{-5}}$ for our $20^\circ\times20^\circ$ ROI, making it a particularly promising target for indirect DM searches.

We have performed a dedicated search for gamma-ray emission from this nearby subhalo candidate using over 17 years of \textit{Fermi}-LAT Pass~8 data. Our analysis framework is tailored to the specific challenges posed by such a nearby extended source. We treat the distance uncertainty by constructing different spatial templates that modify both the morphology and integrated $J$-factor, and we incorporate the mass-induced normalization uncertainty as a log-normal nuisance parameter in the likelihood. A tentative excess is observed in the region, however, while statistically noticeable, its spatial morphology and physical origin are challenging to unambiguously characterize. Adopting a conservative approach, we retain the excess without modeling any extra components in our baseline model. Even under this conservative treatment, the exceptionally large $J$-factor of the nearby subhalo enables us to place stringent constraints on the WIMP annihilation cross section. Contingent on the dynamical confirmation of the subhalo's nature, 
these limits are potentially orders of magnitude stronger than those obtained from dwarf spheroidal galaxies and blind subhalo searches, and are comparable to recent nominal constraints for the ultra-faint satellite Ursa Major III, which would be severely relaxed if uncertainties are considered.

Future progress on this system will likely come from both the dynamical and gamma-ray sides. On the dynamical side, additional pulsar timing measurements and complementary stellar-kinematic probes (e.g.\ \textit{Gaia}~\cite{Gaia:2016zol,2023A&A...674A...1G} and \textit{LAMOST}~\cite{2012RAA....12.1197C,2012RAA....12..723Z} data in the same volume) can refine or challenge the subhalo interpretation of the gravitational perturbation~\cite{2025arXiv251209989P}. On the gamma-ray side, improved models of the Galactic diffuse emission and observations with higher-sensitivity instruments will be essential to clarify the nature of the residual excess and to further tighten indirect DM constraints. Together, these efforts will help determine whether the nearby perturber is indeed a dark matter subhalo or a manifestation of more complex Galactic structure, and will further establish gravitationally selected subhalos as a promising new class of targets for indirect dark matter searches.

\begin{acknowledgments}
Y.L. and X.H. are supported by the National Key Research and Development Program of China (2022YFF0503304). X.H. is supported by the National Natural Science Foundation of China (No. 12322302), and the Project for Young Scientists in Basic Research of Chinese Academy of Sciences (No. YSBR-061).

\end{acknowledgments}

\bibliographystyle{apsrev0}

\bibliography{ref.bib}

\widetext

\end{document}